# Federated AI for building AI Solutions across Multiple Agencies*

**Dinesh Verma**
IBM T J Watson Research Center
Yorktown Heights, NY, U.S.A.
dverma@us.ibm.com

**Simon Julier**
University College London
London, UK
s.julier@ucl.ac.uk

**Greg Cirincione**
Army Research Labs
Adelphi, MD, U.S.A.
gregory.h.cirincione.civ@mail.mil

## Abstract

The different sets of regulations existing for different agencies within the government make the task of creating AI enabled solutions in government difficult. Regulatory restrictions inhibit sharing of data across different agencies, which could be a significant impediment to training AI models. We discuss the challenges that exist in environments where data cannot be freely shared and assess technologies which can be used to work around these challenges. We present results on building AI models using the concept of federated AI, which allows creation of models without moving the training data around.

## 1 Introduction

The use of AI to create data-driven applications holds the promise of revolutionizing how government business is conducted. In this paper, we focus on the sub-field of AI which is based on machine learning.

AI has been used to create applications which include examining images, analyzing sounds, converting speech to text, identifying fraudulent behavior, tracking people from images, detecting fraud and predicting the growth of cities. In its most common approach, an AI-based solution can be viewed as operating in two distinct steps of learning and inference, although there are variations which would allow learning and inference to proceed in parallel.

In the learning stage, training data is used to build an AI model. The AI model captures the patterns and relationships that exist in the training data. There are many alternative ways to capture these patterns, which range from the use of shallow models, e.g. decision trees, inference rules, clusters for classification to deep models such as neural networks and different variations of the neural networks, e.g. convolutional neural networks, or recurrent neural networks. In an abstracted manner, the model could be for finding anomalies, predict the future value of a function or classify the incoming data into two or more categories. During the inference stage, the trained AI model is used to examine new data to reach a conclusion, such as whether the input is anomalous, predict an output, or to determine a category.

The wide-range of AI applications means that they can be meaningfully applied in government contexts. However, there are also several challenges inherent in government contexts, with many of them dealing with the regulations and prohibitions on the data that is maintained within different agencies. This is the problem that we consider in this paper and discuss the approaches that can be used to address these limitations.

The rest of this paper is structured as follows: We discuss the challenges associated with limited availability of data in Section 2. In Section 3, we discuss the concept of federated AI, an approach that allows agencies to share models and model parameters instead of sharing data. We present the results of federated AI on some sample datasets in Section 4. Finally, we draw our conclusions and identify areas for future research.

## 2 Challenges of Data in Federated Agencies

One of the major determinants of effectiveness of any AI based solution is the quality of training data that is available for the task of model building. Since the model is geared towards capturing patterns and relationships in the data, it can only operate well if most patterns and relationships are present in the training data. In general, the larger the amount of training data that is available, the better the quality of AI model that will be available to do the inference part of the solution. This aspect of machine learning is exhibited by the

* This research was sponsored by the U.S. Army Research Laboratory and the U.K. Ministry of Defence under Agreement Number W911NF-16-3-0001. The views and conclusions contained in this document are those of the authors and should not be interpreted as representing the official policies, either expressed or implied, of the U.S. Army Research Laboratory, the U.S. Government, the U.K. Ministry of Defence or the U.K. Government. The U.S. and U.K. Governments are authorized to reproduce and distribute reprints for Government purposes notwithstanding any copyright notation hereon.

learning curve characteristics of different algorithm [Cortes *et al.,* 1994], [Perlich *et al.,* 2003].

In order to build any AI enabled solution, the availability of more data is usually better for creating a model with better accuracy, one that is more likely to capture more patterns. In order to create any model, the required training data may be available with different agencies. Unfortunately, obtaining the right training data for a specific problem in very difficult, and usually only part of the data required to address the problem may be available within a single agency. In order to create the integrated model, one would need to be able to combine and create the data that is available from various sources, and in many cases the relevant data may only be available from another agency.

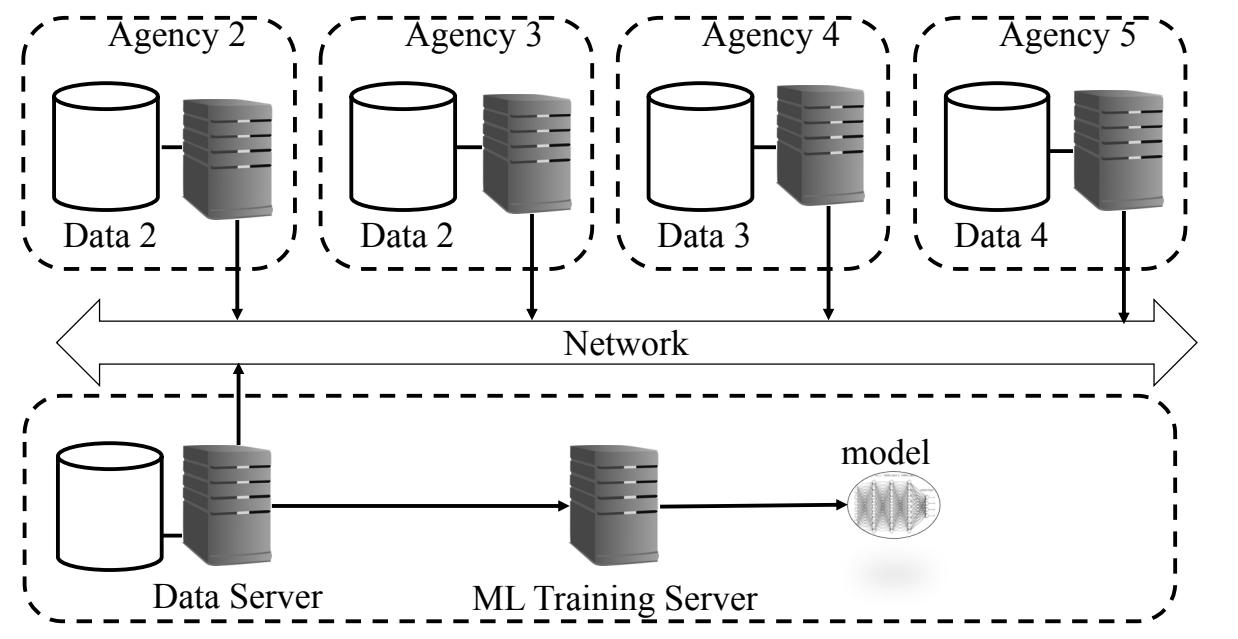

Figure 1. Situation with Inter Agency AI

The general situation which we are trying to address can be seen in Figure 1. One of the agencies, referenced in this paper as Agency 1, is trying to build an AI model and it has access to some training data which is under its own control. It would however benefit from access to a larger set of training data that resides in other agencies (referenced as Agencies 2, 3, 4 etc.). The training data for each agency is accessed via a data server which controls access to the data and may transform the data before providing it to the requesting parties. While it would be useful for the agencies to have free access to all available set of data, it is frequently hard to get access to inter-agency data for a variety of reasons. These include:

*Large Data Size*: The size of data collected by the different agencies can be very large. For example, agencies that collect network data may analyze and store terabytes of data per day. Even agencies that collect structured records about people and personnel may have data collected for hundreds of millions of individuals. The large data size can cause a challenge in moving data to a central location for the purpose of mining and building the appropriate data model.

*Limited Network*: The networks that interconnect different agencies in the government tend to be low bandwidth, usually in comparison to the size of the data that can be stored within each of the agencies. If the data is being collected in environments with limited connectivity, e.g. in remote areas covered only by satellite networks, collection of data into a central location for training may not be viable.

*Regulations*: There may be regulatory prohibitions on the type and amount of data that can be shared among different agencies. Some types of data in an agency may provide valuable information but may be restricted from sharing with other agencies for reasons of privacy, constitutional rights, etc.

*Inter-Agency Trust*: There may be a lack of trust in how the data is handled or managed among different agencies. While different agencies in any government organization cooperated with each other, the trust and cooperation may not be complete. As a result, some agencies may be hesitant to share data with the agency that is building the machine learning model.

*Data Quality*: Different agencies may manage the data with different level of quality or fidelity, as far as the task of curating training data is concerned. While data may be available from many different agencies, the agencies may have collected the data for different purposes, so the way in which the data is maintained and stored may be very different. The data may not be in the format required by the agency, or it may have a different labeling approach.

## 3 Federated AI for Inter-Agency AI

If we can move data from the different agencies to Agency 1 which is creating the machine learning model, then Agency 1 can inject policy-based mechanisms [Agrawal et. al. 2008] to deal with the differences in data format and quality. This would create a centralized policy-based mechanism for collecting training data and create models on the aggregated trained data.

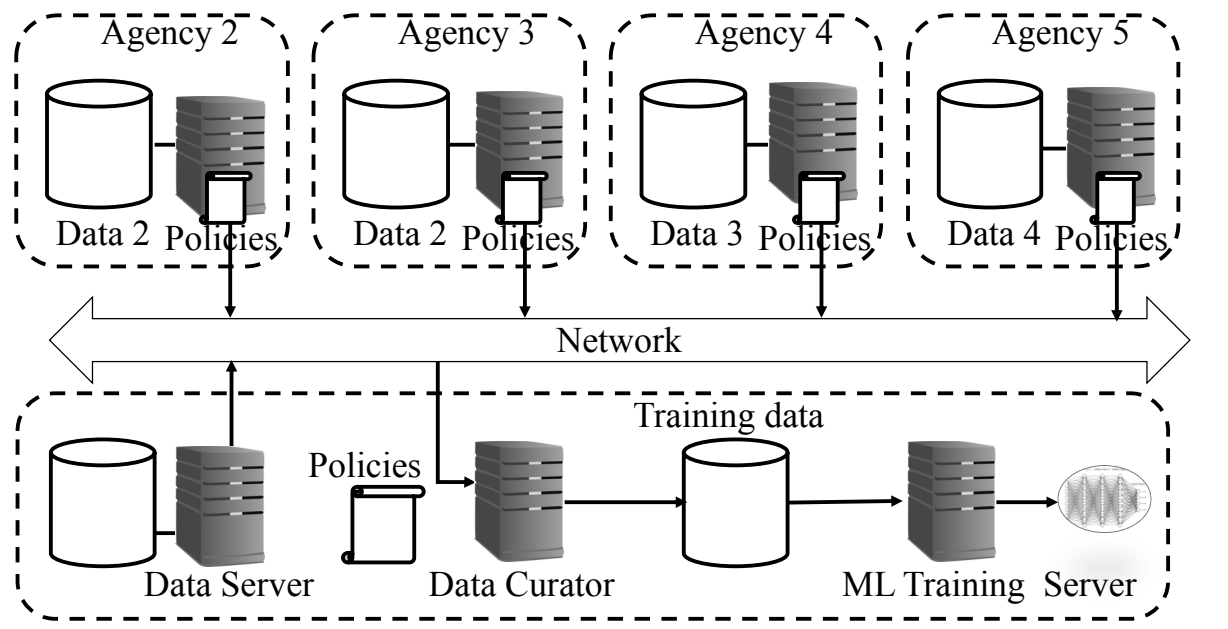

Figure 2. Policy based Centralized Training

On the other hand, if we cannot move data between agencies, an alternative is to create models independently at each of the agencies, and to move models between locations instead of moving data. The models can be moved by simply sharing their parameter values instead of trying to move the data between different sites. We refer to this latter approach as federated AI.

A Centralized Policy based mechanism for creating the federated model would result in the overall architecture as shown in Figure 2. A data curator in Agency 1 interacts with the agencies to collect the training data available from them. Each of the other agencies can also use local policies to decide what data can be delivered to the curator of Agency 1. This allows each agency to have control over how data within their control is sent. Agency 1 would use its curation policies would determine how the data obtained from each agency should be transformed to accept the information. Such transformations could include relabeling, or filtering, rejecting any training data that appears to be erroneous, or applying some transformations on the data that is received. The net result of the policy based curation is the warehousing of all the collected data at one single location, which can subsequently be used to train the model for machine learning.

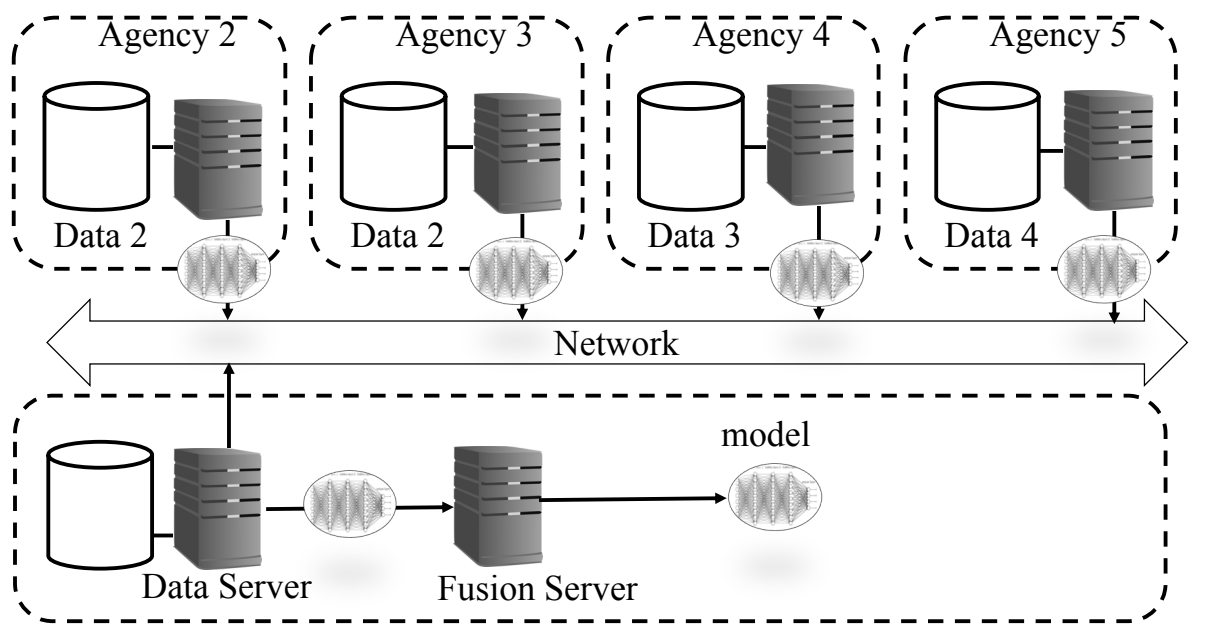

Figure 3. Federated AI

In federated AI, the approach is turned the other way, and data is not moved. Instead of moving data, model parameters are moved so that each agency can help another agency create a training model that is suitable for its purposes. The setup needed for federated AI is as shown in Figure 3. Each agency trains a private model, and exchanges model parameters with Agency 1. Eventually, Agency 1 would end up creating a model that is trained over all the data maintained in all of the agencies.

The choice between the two approaches depends on the trust relationship, compute capacity, network connectivity specifications, and level of cooperation among the different agencies. In the policy-based centralized AI training model, there needs to be sufficient trust between agencies to share that data (possibly with some transformations) to Agency 1. It also needs to have good network bandwidth to transfer the data to Agency 1, but requires little computation support on its own side. On the other hand, federated AI requires sufficient trust among the agencies to run model building on behalf of Agency 1, or to share pre-trained models with Agency 1. It can work with limited network connectivity but requires more computational capacity from each of the agencies. Thus, the choice between centralized AI and federated AI can be viewed as a trade-off between computation cycles and network bandwidth.

One can examine the performance of the centralized approach versus the federated approach as the relative efficiency of network versus compute changes. Suppose it takes $K_n$ units of time to transfer a given size of data across the network to agency one, while it takes $K_s$ units of time to train a machine learning model on the same size of data. The value $N = K_n/K_s$ can be used as the definition of the relative performance of the network compared to computational capacity. The other factors that impact the time taken to train a model would be the number of agencies involved ($A$), and the size of a machine learning model compared to the data that it is trained on, which we can call the model reduction ratio ($M_r$).

With the above definitions, the ratio of the time it takes to train a model using federated AI compared to the time it takes using centralized policy-based learning when each agency has an equal amount of training data can be shown to be:

$$(1+ K_n/K_s)/(1/A + M_rK_n/K_s).$$

The gain in federated learning performance as a function of relative network/compute performance (N) is shown in Figure 4. The graph shows the ratio in the amount of time it takes to train an AI model using federated learning compared to the time it takes compared to the policy-based centralized approach, when each agency has identical computation capacity. The different curves show the ratio for different number of agencies that are involved and assumes that data is split evenly across the different agencies.

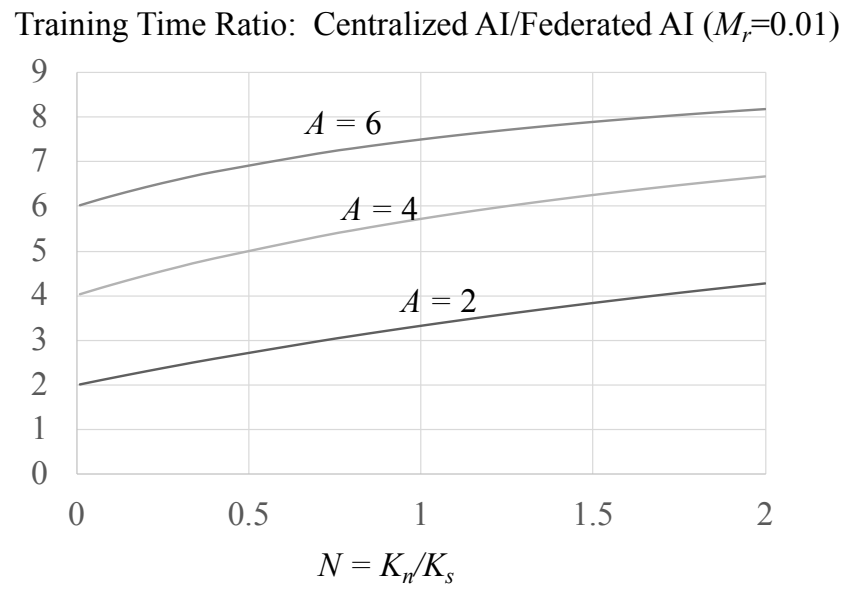

Figure 4. Relative Performance of the two approaches

As apparent in Figure 4, federated AI would usually perform better compared to centralized approach, even for relatively low values of $N$, which indicates the performance of the network is very high. As the network performance becomes worse, (N becomes high) with a high latency network, the benefits of federated AI become more pronounced. Centralized AI would have better performance only when Agency 1 has special types of servers which make the computation capacity relatively fast for the central learning while requiring a very large time for training at the other agencies.

In the asymptotic case, where the model size relative to the size of the data can be ignored, the performance gain of federated AI will be characterized by $A(\mathrm{I}+ K_n/K_s)$.

Although federated AI would out-perform centralized AI in most cases, there may be scenarios where centralized AI would still be the preferred approach. The centralized AI approach requires less dependency on the other agencies from Agency 1's perspective since they only need to provide access to their data. Furthermore, policies can be made to deal with the nuances of different data formats or data quality across agencies.

While Federated AI has a clear win over centralized AI as far as performance measured in the model training time is concerned, it does require more computational capability support from the other agencies. As a result, different flavors of federated AI may need to be developed depending on the level of cooperation that may exist among the different agencies.

## 4 Flavors of Federated AI

The flavors of federated AI reflect the variations on the approach that needs to be developed based on the truest relationship among different agencies. In defining each of the flavors, we assume an agent-based approach for federated AI. The federation process works on having an "*federation client*" that is executed by each of the agencies. This client can be downloaded and executed from a location hosted by agency 1. The overall approach is shown in Figure 5. The federation clients in each agency interact with a federation server in agency 1.

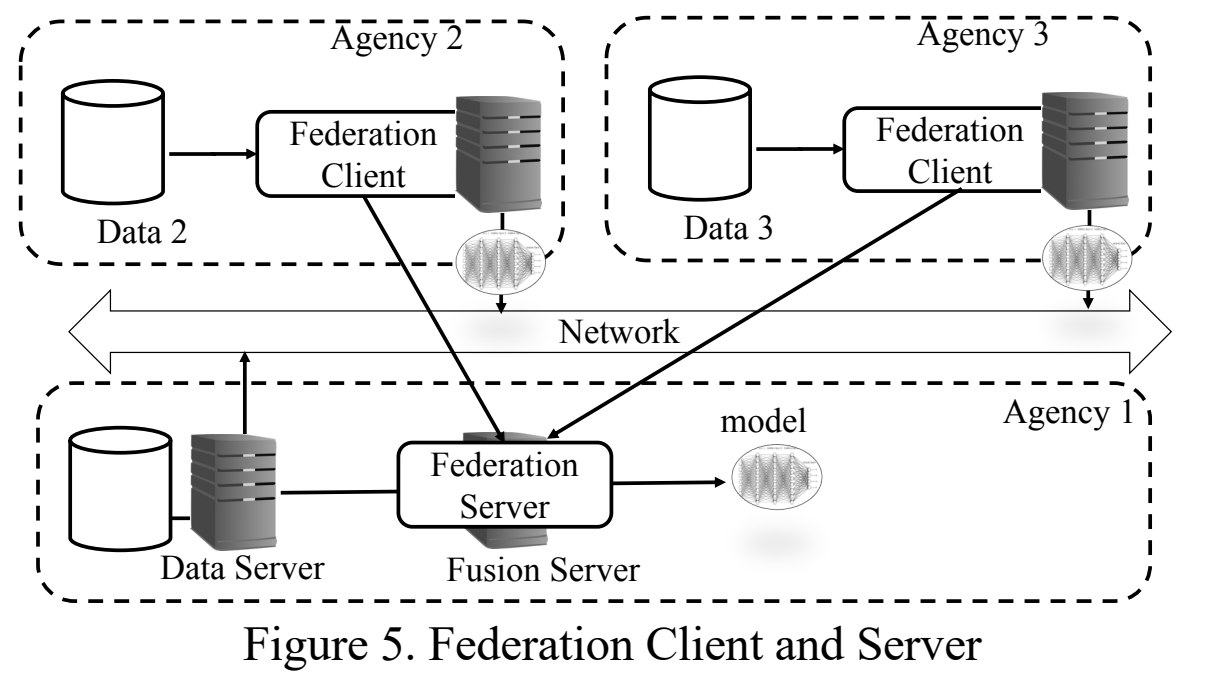

Figure 5. Federation Client and Server

Each agency is willing to run the federation client to help out agency 1 in building its training model. Even as the agencies cooperate, they may be placing various kinds of constraints on what the federation client does, and when the federation client may run. Some of these constraints include:

- Agency 1 can control the time of execution of federation client, and enable them to access the data at other agencies at periodic intervals. This would allow agency 1 to execute all the federation clients at the same time interacting with the federation server, incrementally building the model that is required.
- Agency 1 cannot control how the federation clients communicate with the federation server. Each federation client, however, can train a local model completely, and share that model with the federation server.
- The other agencies are only willing to provide Agency 1 with a fully trained model. Agency 1 federation client is limited to transmitting this pre-trained model to the federation server.

Each of these constraints would result in different flavors of federated AI. In this section, we look at some of these flavors and see how the accuracy of models trained using the federated approach compares to that trained using the central approach.

### 4.1 Flavor 1: Synchronized Online Federation

In this flavor of federated learning, the federation client at different agencies can be scheduled to run at the same time as the federation server. The federation server decides on the type of model that is to be trained by each of the federation clients. Each federation client will use the same type of AI model with the same hyper-parameters for its training. Each federation client trains the model parameters on a small mini-batch of local training data, sends the parameters to the federation server, which has the task of combining and averaging all of the parameters together. In the next training round, each agent takes the averaged parameters and trains its model with that as the average approach.

This synchronized federation can work on machine learning models that use synchronized gradient descent [McMahan *et. al.* 2017] with an additive loss function, i.e. the measure of error from the learnt model against given data points. In effect, each agent computes the model parameters to best fit their data, and the loss from the different agents can be averaged at the federation server to determine the right parameters. This approach would work for the majority of neural networks.

In order to train a model that captures the relationship among the different split data properly, the system needs to go through multiple rounds of synchronization. With increasing rounds of synchronization, the performance of the model approaches that of collecting all the data in a central location and then training the model.

Figure 6 shows the accuracy of this synchronization approach on the MNIST data set [Lecun *et. al.* 1998] when it is split randomly into ten agencies. The agencies are each training a convolutional neural network with two convolutional layers using rectified linear unit activation followed by two fully connected layers, the first one using rectified linear unit activation and the second one using softmax to classify the digits. The accuracy of this neural network on centralized MNIST data is 98.8%. As the number of synchronization points among the different agents are increased, the accuracy of the federated AI approaches that of the centralized AI approach. This observation is consistent with the results shown in [McMahan *et. al.* 2017] and [Wang *et. al.* 2018].

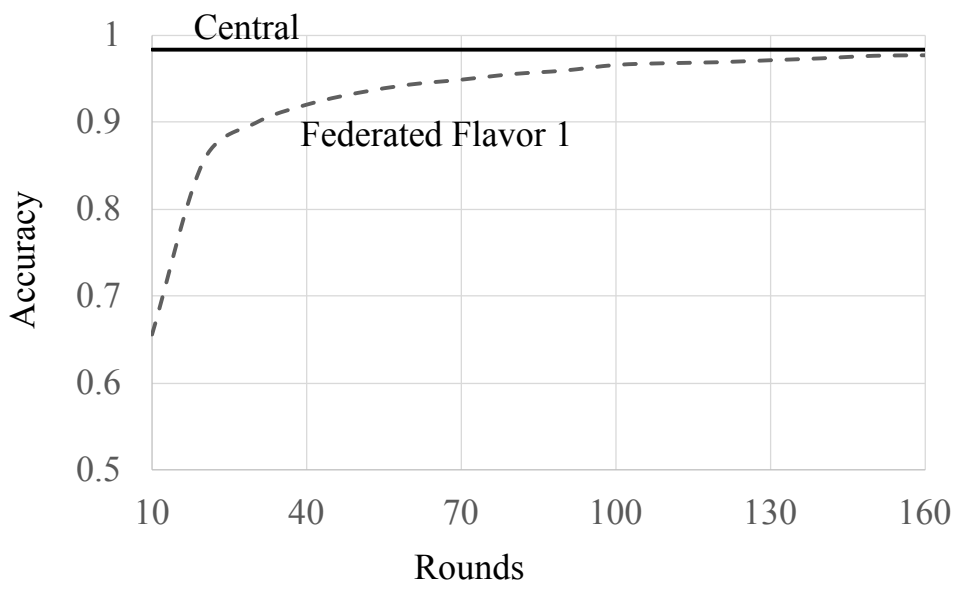

Figure 6. Accuracy of Federation AI Flavor 1

## 4.2 Flavor 2: Unsynchronized Model Movement

In this flavor of federated learning, the federation clients are not allowed to synchronize their model building process over every mini-batch. Instead, each agency can train the data over the local data set that it has access to. This situation may arise in environments where the other agency may not be willing to open its local systems to download and run a federation client, but is willing to take a model specification, and run it locally on data that is available locally.

In this flavor of the federated learning, the federation server in agency 1 has to take the model that it has from its local data, and then send the models over to each agency for training further using the data set at the agency. In effect the model is moved across all of the agencies.

Figure 7 shows the resulting accuracy of the model as it goes through the model movement across all the different data sets. The model migration approach works fairly well for this particular data set and model, and at the end of having traversed all of the local data instances, the model has the accuracy comparable to that of the centrally trained model.

Note that the horizontal axis of Figure 7 reflects the number of peer agencies that the model has been moved to. In this flavor of federated AI, each model is trained through all the data stored at an agency together, as opposed to the concept of synchronization rounds in flavor 1 where each round uses data from each of the different agencies. The bottom axis, therefore, is different than the one shown in Figure 6. For the MNIST data set, each agency data reflects about 46 rounds of synchronization using the flavor 1 approach. Thus, the initial

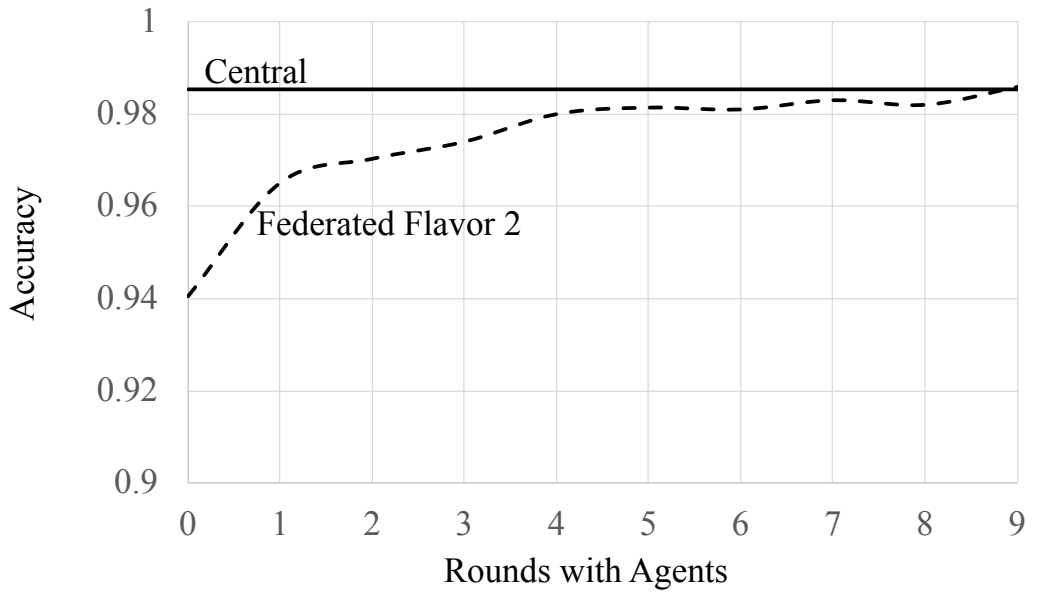

Figure 7. Accuracy of Federation AI Flavor 2

accuracy starts out comparable to the accuracy at round step 50 of the flavor 1 model.

## 4.3 Flavor 3: Limited Data Exchange Approach

The previously described two flavors work well when the data across different agencies is randomly split. However,

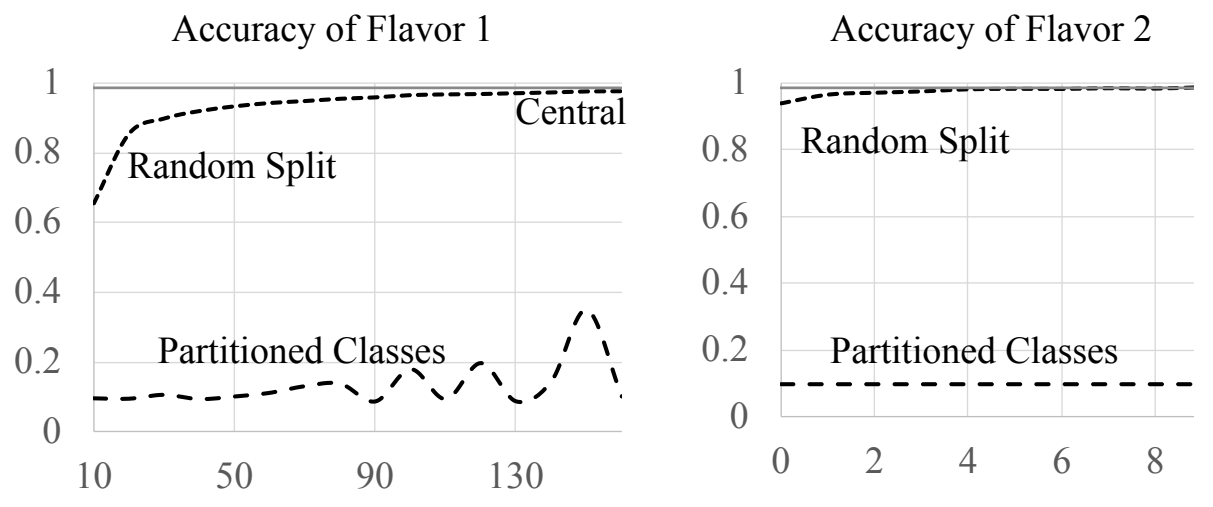

Figure 8. Problems with Partitioned Classes

when the data is skewed, and different classes have different types of data, then two flavors of federated AI do not perform as well.

Figure 8 shows the accuracy of the two flavors of federated AI when the data among different agencies is partitioned so that data for a specific class belongs only to one of the agencies. In this particular case, we have taken the MNIST data set and split it among 10 agencies so that each agency only has one class of data. Neither of the two flavors for creating an AI model works well in this instance. The flavor 1 approach, which resulted in a very good partition with random split of data, ends up with some very poor results. While the system seems to be improving its accuracy relatively slowly with the increasing rounds of synchronization, the growth in accuracy is anemic compared to the case where the data was randomly split.

The flavor 2 of federated learning performs even worse, with the accuracy being static regardless of the number of agencies from which the daa is being trained. An examination of the details of what is happening in the second flavor indicates the problem that arises in this approach. After each round of data training, the AI model is trained to predict all input data as belonging to the class for which it was trained most recently. With the test data comprising of equal set among the ten classes, the resulting model is able to only predict 10% of the classes. The problem is that the neural network training algorithms are biased towards the latest set of data it is being trained, and splitting the data in a skewed manner confuses them to recognize the latest class preferentially.

This effect can be mitigated if we are able to provide the machine learning model at each of the stages with a little bit of data from the other classes. The model is now able to account for existence of other types of classes, and does not get biased too heavily with the class it has been seeing in the latest reincarnation.

The results from the training process with a limited amount of data exchange is shown in Figure 9. The left hand side shows data exchange added to flavor 1 of federated learning, while the right hand side shows data exchange added to flavor 2 of federated learning.

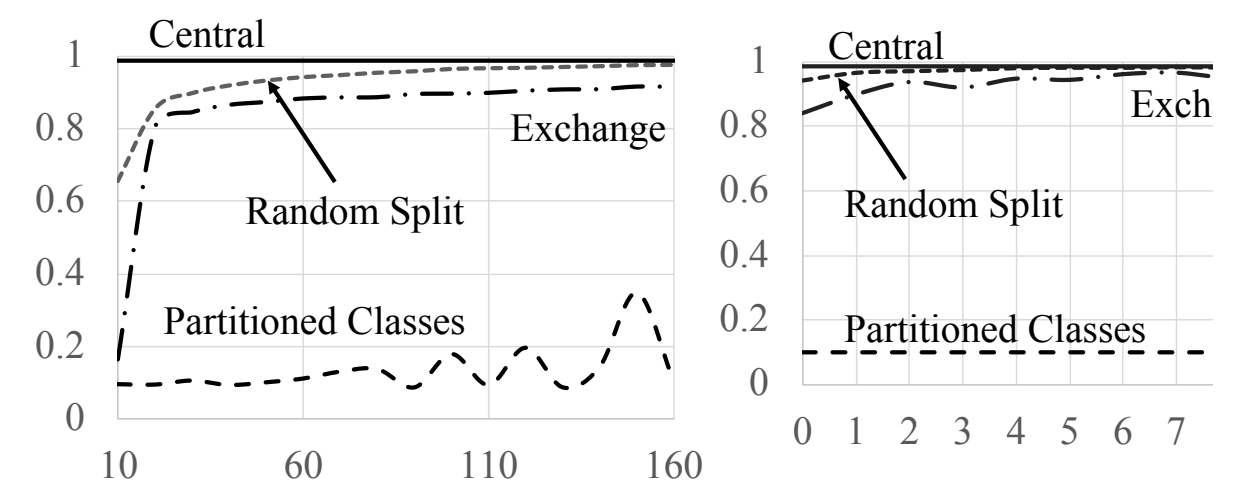

Figure 9. Accuracy of Federation AI Flavor 3

The data exchange in Figure 9 reflects an exchange of 128 samples per class to each agent. This reflects a transfer of less than 2% of the data that is for each class (for MNIST, it is 6,000 points per class). Even with a small amount of data exchanged, the accuracy of the model increases significantly, approaching the accuracy that was attained with the simpler previous flavors of the algorithm when the data was split randomly. Comparing the results to Figure 8, the advantages of a limited data exchange are obvious.

A natural question in this regard is the impact of the amount of data exchange and its impact on the accuracy of the resulting model. Figure 10 shows the data exchange approach added to flavor 1. The simulation runs for 50 rounds of exchanges, and compares the increase in the accuracy of the resulting model. From the shape of the results, even a small amount of data exchange (~1%) can result in significant improvement of accuracy in building the different models.

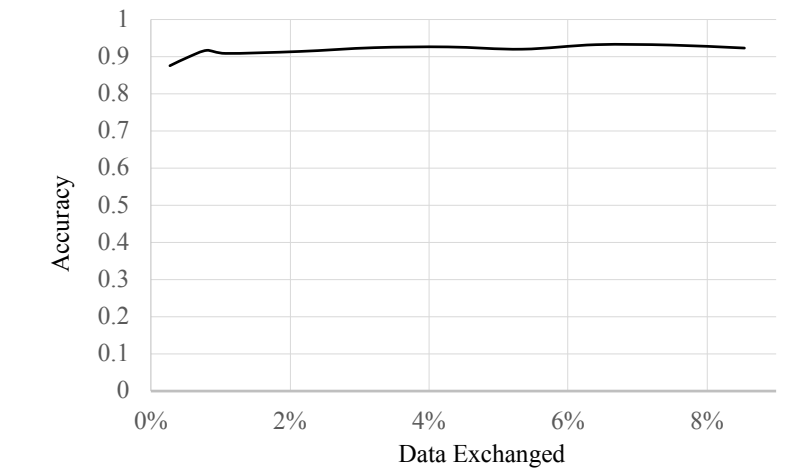

Figure 10. Data Exchange impact on Accuracy of Federation AI Flavor 3

The net resulting accuracy is sufficiently close to that of centralized training approach, and can be improved further if the number of rounds are increased beyond 50, extrapolating from the trend shown in the results of Figure 9.

### 4.3 Other Flavors

The previously described flavors have the focus on making models with good accuracy metrics. However, Government agencies may have other concerns, such as security of data, potential leakage of model parameters to other agencies, or maintaining provenance of models and training data.

Each of the flavors described earlier in the section can be augmented with cryptographical methods to answer those requirements. The model building process can be restructured so that partial homomorphic encryption techniques will allow the model building in each of the other flavors to occur, without necessarily sharing information content or model parameters with each other. At the cost of performance degradation, fully homomorphic schemes can also be used.

Practical solutions may often require several AI models which are chained together to create a complete end to end solution. Those solutions would need additional flavors to support these chains. Similarly, provenance can be added as part of the information exchange among agencies.

While we are not describing these other flavors in this paper, prototypes implementing those functions can be added relatively easily to the three flavors described earlier.

## 4 Related Work

The challenges of creating machine learning models across several agencies is related to the problem of learning models across multiple coalition partners in a coalition environment. The challenges of distributed learning in coalition environments are described by [Verma and Julier 2017], and many of the issues carry over to the challenge of training models on data across agencies of the same government.

The first flavor of federated AI can be viewed as an application of the approach for handling decentralized data described by [McMahan *et. al.* 2017], although they did not consider the impact of partitioned classes in training data. The approach used in flavor 1 was analyzed by [Wang *et. al.* 2018] to optimize the intervals at which models are synchronized. An alternative approach for combining models trained on distributed data using boosting and ensemble methods [Rokach 2010] was proposed by [Verma et. al. 2018] but its performance was poor compared to the flavors proposed in this paper.

While this paper has focused on combining neural network models, there has been work on combining models of other types. Approaches for combining decision trees were reviewed by [Strecht 2015]. Models that can be expressed using their Fourier transforms can be combined using approaches proposed by [Karagupta and Park 2004].

Another approach to solve the problem of data across agencies is to move data after anonymization to preserve privacy [Hua and Pei 2008]. Such transformations may augment the centralized policy-based mechanisms described earlier in this paper.

## 5 Summary and Conclusions

In this paper, we have discussed the challenges associated with the training of AI models where the training data set is distributed across multiple agencies, and data cannot be moved across organizations easily. We have examined three different flavors of federated AI, and discussed how a small amount of data exchange can significantly improve the performance of federated AI.

While the three flavors of federated AI provide a set of solutions to some common issues preventing sharing of data, they are only the starting point of an approach which addresses the concerns of disparate training data spread across different agencies. While some of those approaches can be addressed using policy-based mechanisms, other challenges would require new flavors of federated AI.

As an example of the challenges that still need to be addressed, in some cases the agency may find it more convenient to run the federation server on a Government hosted cloud. If an agency is not comfortable sharing the model parameters to a cloud-based servers, techniques that do not reveal the model parameters to the federation servers would need to be developed.